\documentclass[journal=aamick,manuscript=article]{achemso}

\usepackage{epsfig}
\usepackage[]{graphicx}
\usepackage{bm}
\usepackage{amssymb}
\usepackage{chemformula} 
\usepackage[T1]{fontenc}

\author{L. J. Stanley}

\affiliation{National High Magnetic Field Laboratory, Florida State University, Tallahassee, FL 32310, USA}
\alsoaffiliation{Department of Physics, Florida State University, Tallahassee, FL 32306, USA}

\author{ Hsun-Jen  Chuang}
\altaffiliation{Present address: Nova Research, Inc., Alexandria, VA 22308, USA}
\author{Zhixian Zhou}
\email{zxzhou@wayne.edu}
\affiliation{ Physics and Astronomy Dept., Wayne State University, Detroit, MI 48202, USA}
\author{M. Koehler}
\affiliation{Dept. of Materials Science and Engineering, The University of Tennessee, Knoxville, TN 37996, USA}
\author{J. Yan}
\affiliation{Dept. of Materials Science and Engineering, The University of Tennessee, Knoxville, TN 37996, USA}
\alsoaffiliation{Oak Ridge National Lab, Oak Ridge, TN 37830, USA}
\author{D. Mandrus}
\affiliation{Dept. of Materials Science and Engineering, The University of Tennessee, Knoxville, TN 37996, USA}
\alsoaffiliation{Oak Ridge National Lab, Oak Ridge, TN 37830, USA}
\author{Dragana Popovi{\'c}}
\email{dragana@magnet.fsu.edu}
\affiliation{National High Magnetic Field Laboratory, Florida State University, Tallahassee, FL 32310, USA}
\alsoaffiliation{Department of Physics, Florida State University, Tallahassee, FL 32306, USA}

\title[Low-Temperature 2D/2D Ohmic Contacts in WSe$_2$ FETs]{Low-Temperature 2D/2D Ohmic Contacts in WSe$_2$ Field-Effect Transistors as a Platform for the 2D Metal-Insulator Transition}

\abbreviations{hBN, 2D, MIT, MOSFETs, TMDs, SB, EBL}

\begin{document}

\begin{abstract}

We report the fabrication of hexagonal-boron-nitride (hBN) encapsulated multi-terminal WSe$_2$ Hall bars with 2D/2D low-temperature Ohmic contacts as a platform for investigating the two-dimensional (2D) metal-insulator transition. We demonstrate that the WSe$_2$ devices exhibit Ohmic behavior down to 0.25 K and at low enough excitation voltages to avoid current-heating effects. Additionally, the high-quality hBN-encapsulated WSe$_2$ devices in ideal Hall-bar geometry enable us to accurately determine the carrier density. Measurements of the temperature ($T$) and density ($n_s$) dependence of the conductivity $\sigma(T,n_s)$ demonstrate scaling behavior consistent with a metal-insulator quantum phase transition driven by electron-electron interactions, but where disorder-induced local magnetic moments are also present.  Our findings pave the way for further studies of the fundamental quantum mechanical properties of 2D transition metal dichalcogenides using the same contact engineering.

\end{abstract}

\section{Introduction}

Materials that can be tuned easily from the limit of good metals to poor conductors and even insulators continue to be of great interest both for their scientific and technological importance.  Indeed, the best known example is probably the Si metal-oxide-semiconductor field-effect transistor (MOSFET), which forms the basis of modern semiconductor technology and in which the conducting properties can be tuned by applying an external electric field on the gate.  In general, however, the microscopic origin of the metal-insulator transition (MIT) in most materials of interest is not well understood.  Furthermore, many novel materials behave effectively as two-dimensional (2D) systems, but the 2D MIT remains one of the most fundamental open problems in condensed matter science \cite{Abrahams2001, Kravchenko2004, Spivak2010, Dobrosavljevic2011,DP-CIQPT,Shashkin-chapter2017,Dragana2017}.  In fact, the 2D electron system in Si MOSFETs is the only system in which detailed and comprehensive investigations of this quantum (i.e. $T=0$) phase transition have been carried out so far.  Based on Si MOSFET studies, it has been suggested that there are three universality classes for the 2D MIT \cite{Dragana2017,Lin2015}.  However, to test the universality of the observed behaviors, it is necessary to probe 2D systems in other types of materials, in particular beyond conventional semiconductor heterostructures.  Thus 2D systems in novel 2D-semiconductor-based transistors represent specially promising candidates for such studies.  However, the efforts to study the 2D MIT in 2D semiconductors such as transition metal dichalcogenides (TMDs) have been limited \cite{nihar2015,moon_soft_2018}, especially at low temperatures, because of the high-resistance, non-Ohmic contacts. 

The primary cause of the high contact-resistance and non-Ohmic contacts is the tendency for 2D semiconductors to form a substantial Schottky barrier (SB) with most contact metals~\cite{Fang2012, Das2013, Allain2015, Das2013a, Liu2014}. As temperature decreases, the suppression of thermal emission and thermally assisted tunneling current over the SB leads to a drastic increase of contact resistance and increasingly non-Ohmic behavior. 

In Si MOSFETs, low-temperature Ohmic contacts are achieved via degenerately doping the contact regions to reduce the SB width and therefore enable efficient quantum mechanical tunneling of carriers through the metal-semiconductor interface. This is realized by selective ion-implantation doping of the drain and source contact regions under the metal electrodes. However, the atomic thickness of strongly layered 2D materials, such as TMDs, prevents controlled and spatially defined heavy doping by ion implantation. Therefore, it is essential to minimize the SB height in order to achieve low-temperature Ohmic contacts in 2D semiconductor devices. Metals of very low and high work functions have been used in attempts to  eliminate the SB for the electron and hole channels of 2D semiconductors, respectively, but yielded only limited success largely due to the Fermi-level pinning effect~\cite{Das2013a, Movva2015}.

Multiple research groups around the world have attempted to eliminate the Fermi-level pinning and eradicate the SB at metal/2D-semiconductor interfaces by experimenting on different fabrication methods. However, their reported results exhibit significant divergence and point to drastically different origins of Fermi-level pinning in nominally similar metal/2D-semiconductor junctions. For example, Xu \textit{et al.} reported low-temperature Ohmic contacts in hBN-encapsulated TMD devices based on selective etching and conventional electron-beam evaporation of metal electrodes~\cite{Xu2016}.  They claimed charge traps and disorder formed at the TMD/SiO$_2$ interface as a main contributor to Fermi-level pinning. On the other hand, Liu \textit{et al.} attributed the Fermi-level pinning effect at metal/MoS$_2$ contacts mainly to the structural and chemical defects created during the metal deposition~\cite{Liu2018}.  Additionally, the recent work by Zheng \textit{et al.} points to resist contamination at the metal/2D-semiconductor interfaces as a dominant source of Fermi-level pinning~\cite{Zheng2019}. Furthermore, the metal/2D-semiconductor contact quality intricately depends on the processing environment such as the deposition-chamber vacuum, deposition rate, and metal topography, making it challenging to produce devices with consistent, stable performance~\cite{Movva2015, Xu2016, English2016}. 

Various alternative contact-engineering techniques aimed to eliminate/minimize the SB, such as surface and substitutional doping~\cite{Fang2012,Feng2013, Kiriya2014, Yang2014, Suh2014}, phase-engineering~\cite{Kappera2014, Cho2015}, use of graphene contacts~\cite{Chuang2014, Roy2014, Das2014, Du2014}, insertion of a work-function-matching buffer layer~\cite{Chuang2014a}, and edge contacts~\cite{Cheng2019, Jain2019}, are still deficient as they lack sufficient air or thermal stability, have limited spatial definition of the contact regions, or do not offer true Ohmic contacts at low temperatures. 

Some of us have recently demonstrated a 2D/2D contact strategy for engineering low-temperature Ohmic contacts to a variety of TMDs~\cite{Chuang2016}. In this prior study, Ohmic contacts with very low SB were achieved down to 5 K by van-der-Waals assembly of degenerately doped TMDs as contact materials and undoped TMDs as channel materials. However, the two-contact TMD FET devices used in our prior study were not suitable for the study of the 2D MIT, which requires multi-contact TMD devices shaped in Hall-bar geometry. 

In this study, we have fabricated high-quality multi-contact Hall bars consisting of an hBN-encapsulated WSe$_2$ channel with 2D/2D contacts that are Ohmic down to 0.25~K. We further show that non-Ohmic behavior at all temperatures may be induced by large currents via the trivial heating effect.  By performing four-terminal measurements in the Ohmic regime, we demonstrate the critical behavior of the conductivity $\sigma$ as a function of $n_s$ and $T$, consistent with the existence of a MIT.  Furthermore, our scaling analysis finds that the critical exponents agree with those found in few-layer ReS$_2$ devices \cite{nihar2015} and in relatively clean 2D systems in Si MOSFETs in the presence of local magnetic moments \cite{Feng2001}, one of the proposed universality classes \cite{Lin2015} for the 2D MIT.  Our results, therefore, demonstrate that these high-quality 2D-semiconductor Hall-bar devices with low-temperature Ohmic contacts provide an ideal platform for investigating the 2D MIT beyond Si-based systems.  

\section{Materials and methods}

The Hall-bar devices were fabricated by van der Waals assembly of WSe$_2$ with degenerately doped Nb$_{0.005}$W$_{0.995}$Se$_2$ stacked as the electrical contacts, as described previously \cite{Chuang2016}.  To preserve the intrinsic channel properties, the entire WSe$_2$ channel was fully encapsulated by hBN. After fabricating electrical leads to the degenerately doped Nb$_{0.005}$W$_{0.995}$Se$_2$ contacts by electron beam lithography (EBL) and metal deposition, a second EBL step and reactive ion etching were used to shape the device into a Hall-bar geometry.  The detailed fabrication steps were as follows.

First, 10--30 nm thick hBN flakes exfoliated on degenerately doped Si with 280 nm of thermal oxide were used as ultra-flat and ultra-smooth substrates to minimize dangling bonds and charge traps. Next, a few-layer WSe$_2$ flake was placed on an hBN substrate by a dry transfer method~\cite{Chuang2016, Wang2018, Hong2015}. Subsequently, another hBN flake was stacked on top of the WSe$_2$ channel also by the dry transfer method, serving as the top encapsulation layer. Care was taken to ensure that the lateral size of the top hBN is slightly smaller than that of the WSe$_2$ flake such that the perimeter of the WSe$_2$ flake is exposed.  To fabricate 2D/2D contacts to the WSe$_2$, thin flakes of degenerately p-doped WSe$_2$ were placed around the perimeter of the WSe$_2$ flake sandwiched between a bottom hBN substrate and a top hBN using the same dry transfer method~\cite{Cui2015}. Stacked heterostructure was annealed at 400$^{\circ}$C for 30 minutes under vacuum without forming gas. Then, metal electrodes were fabricated on top of the p-doped TMD contacts by EBL and electron-beam assisted deposition of Ti/Au (5nm/50nm).  Finally, the device was shaped into a Hall-bar geometry by another step of EBL followed by SF$_6$ plasma etching. Thin flakes of intrinsic WSe$_2$, degenerately p-doped WSe$_2$ and hBN were all produced by mechanical exfoliation of bulk crystals and identified using optical microscopy and Park-Systems XE-70 noncontact mode atomic microscopy. Bulk crystals of WSe$_2$ and degenerately p-doped WSe$_2$ were synthesized by chemical vapor transport~\cite{Chuang2016, Mukherjee2017}. 

Mobility measurements of the  WSe$_2$ devices were made by a Keithley 4200 semiconductor parameter analyzer in a Lakeshore Cryogenic probe station under high vacuum (1 x 10$^{-6}$ Torr) or in a Quantum Design PPMS.

All other electrical transport measurements were taken using the standard four-probe ac technique with SR7265 lock-in amplifiers (typically at $\sim 11$~Hz) along with an ITHACO 1211 current preamplifier.  In the regime of high sample resistance, we also used the SR 5113 voltage preamplifier. Measurements were made either in a standard VTI system or under high vacuum (1 x 10$^{-6}$ mbar) in a $^3$He cryostat (Oxford HelioxVL system). 

\section{Results and discussion}

\begin{figure}[!t]
\centering
    \includegraphics[width=\textwidth]{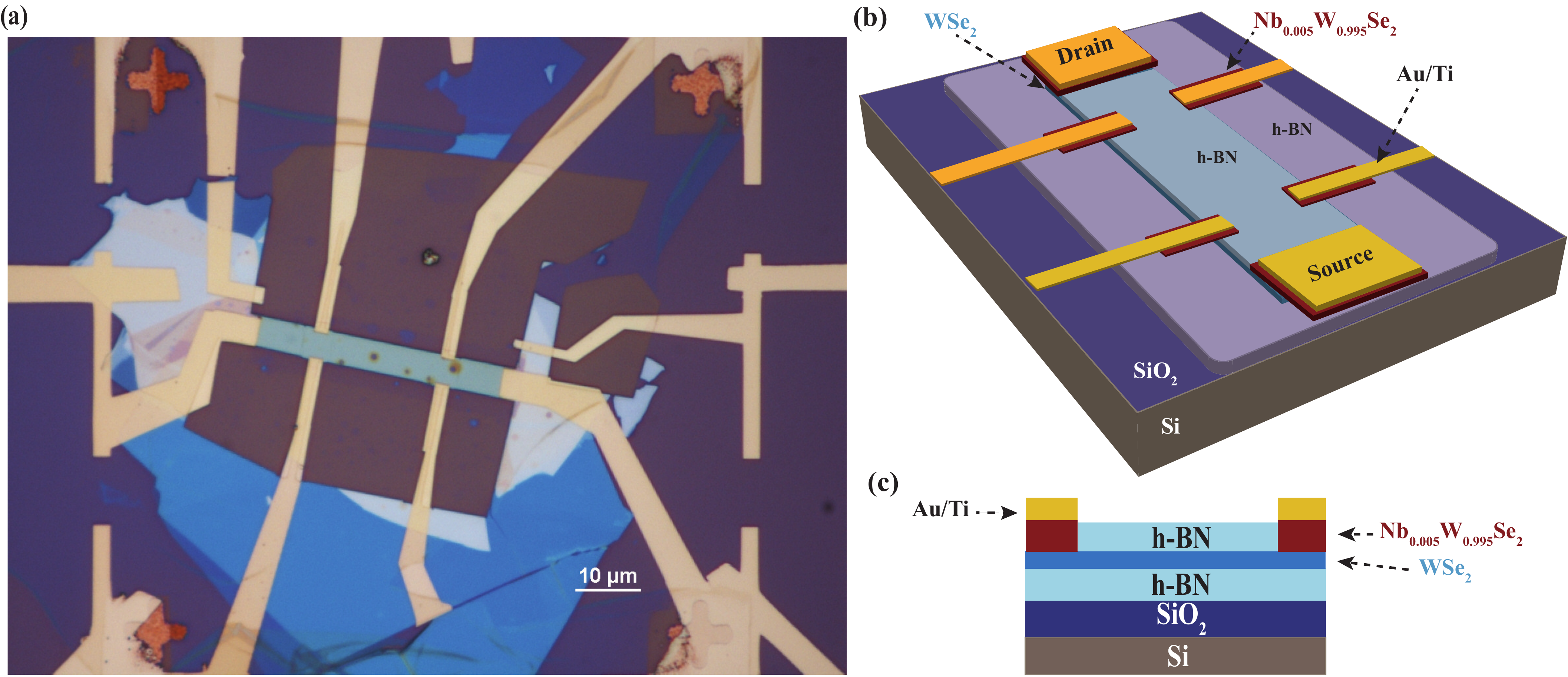}
\caption{(a) Optical micrograph of a representative Hall-bar device (sample No1). (b) A schematic diagram of the WSe$_2$ FET device. (c) Schematic side view of the contact structure.}
\label{fig1}
\end{figure}

  Figure~\ref{fig1}(a) shows one of our WSe$_2$ devices, sample No1, with channel thickness of 8.4~nm (12 layers) and lateral dimensions $L\times W=19.4\times 4.7~\mu$m$^2$ ($L$--length, $W$--width); another sample that was measured down to $\sim0.25$~K, No4, had  channel thickness of 6.5~nm (9 layers) and $L\times W=13.4\times 4.9~\mu$m$^2$. The 2D nature of such few-layer WSe$_2$ has been confirmed by Shubnikov-de Haas oscillations in magnetic fields up to 14 T~\cite{Xu2016}. A schematic diagram of our WSe$_2$ device structure is presented in Figures~\ref{fig1}(b) and \ref{fig1}(c).

\begin{figure}[t]
\centering
    \includegraphics[width=\textwidth]{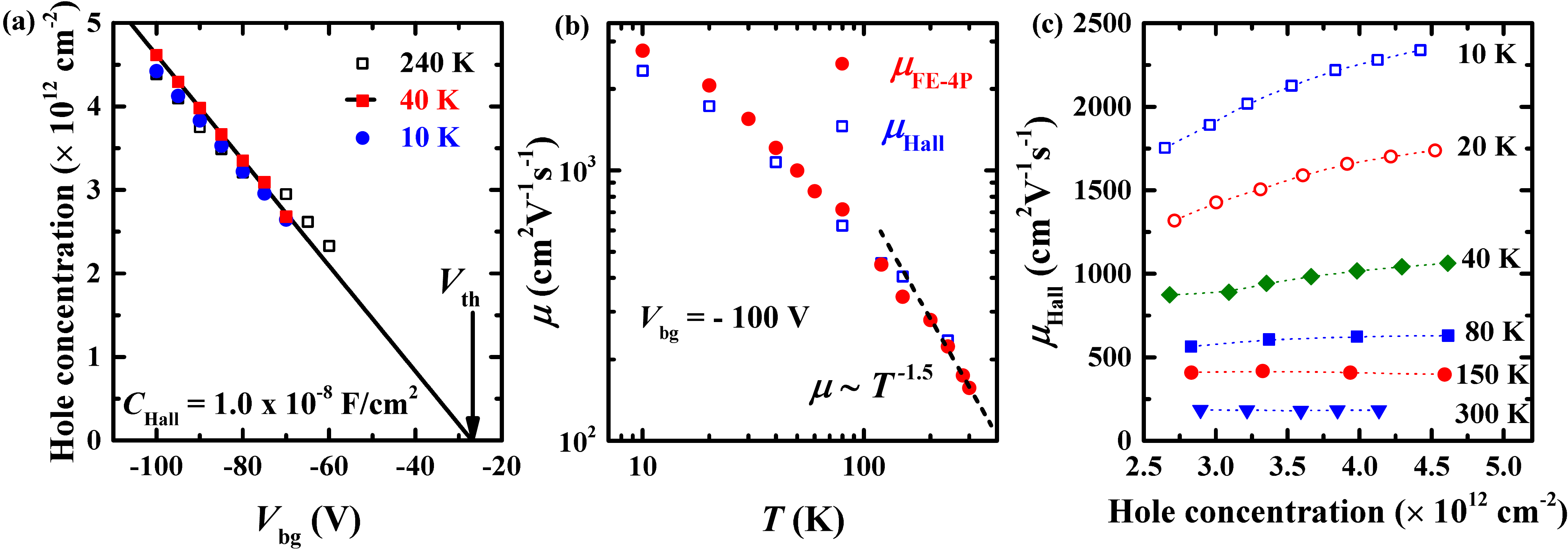} 
      \caption{Sample No1.  
      (a) Hole concentration determined from Hall measurements as a function of back-gate voltage for multiple temperatures. Solid line is a linear fit of the 40 K data, which extrapolates to $V_{\mathrm{th}}\approx -27$~V. (b) Hall and four-probe field-effect mobility as a function of temperature, measured at $ V_{\mathrm{bg}}=-100 $ V and in the high gate-voltage linear region ($-100 < V_{\mathrm{bg}} < -80$ V), respectively. Dashed line is a power-law fit $\mu\sim T^{-\gamma}$ with $ \gamma\approx 1.5 $. (c) Hall mobility as a function of hole density for selected temperatures. Dashed lines guide the eye.  }
  \label{fig2}
\end{figure}

To characterize the quality of our hBN-encapsulated WSe$_2$ devices, we have measured the mobility as a function of temperature and carrier density in three devices (No1, No2, No3) with the WSe$_2$ channel thickness ranging from $ \sim $ 4.9 to 8.4 nm. All three devices exhibited qualitatively and quantitatively similar behavior, and data from a representative device are shown in Figure~\ref{fig2} (results from two other devices are presented in Figure~S1 in the Supporting Information).  In order to reliably determine the mobility, it is important to accurately determine the carrier density, which can be either calculated using a parallel-plate capacitor model or extracted from Hall measurement~\cite{Li2014}. However, it is challenging to accurately determine the carrier density of 2D semiconductors from the parallel capacitor model because the threshold voltage of a 2D FET is often affected by various extrinsic factors such as charge trapping and non-ideal contacts~\cite{Andrews2020}.  Carrier density derived from the Hall measurement may also be affected by artifacts such as non-ideal Hall-bar geometry and disorder at the channel/substrate interface~\cite{Pradhan2015}. These extrinsic effects are eliminated/minimized in our hBN-encapsulated WSe$_2$ devices with ideal Hall-bar geometry and low-temperature Ohmic contacts. As a result, the carrier densities derived from Hall measurement are fully consistent with parallel-capacitor model and also nearly temperature independent, indicating that they represent the true carrier densities induced by the gate voltage. 

As shown in Figure~\ref{fig2}(a), the hole densities determined by Hall measurement, $n_{\mathrm{Hall}}$, vary linearly with the gate voltage and are independent of temperature between 10~K and 300~K.  The slope of the hole density ($n_{\mathrm{Hall}}$) vs. gate voltage ($V_{\mathrm{bg}}$) yields an equivalent back-gate capacitance of $C_\mathrm{Hall} = 1.0$ nF/cm$^2$, which is in excellent agreement with the geometric capacitance $C_\mathrm{Geo}$. Here the geometric capacitance is the equivalent series capacitance of SiO$_2$ and hBN substrate calculated based on the parallel plate capacitor model and a dielectric constant of 3.5 for hBN~\cite{Lee2013, Chamlagain2014}. The excellent agreement between the capacitances determined by the two different methods strongly indicates that the extracted carrier density values represent the true carrier density in the WSe$_2$ channel. For sample No1, the carrier density $n_s$ can be found using $V_{\mathrm{bg}}$: $ n_s= C_\mathrm{Hall}\vert V_{\mathrm{bg}}-V_{\mathrm{th}}\vert /e$; $V_{\mathrm{th}} \approx -27 $ V. This gives $n_s=0.62 \vert V_{\mathrm{bg}}+27\vert ~(10^{11}$ cm$^{-2})$. 

Figure~\ref{fig2}(b) shows the temperature dependence of both Hall mobilities [$\mu_{\mathrm{Hall}}=\sigma/(en_{\mathrm{Hall}})$] measured at the highest back-gate voltage ($ V_{\mathrm{bg}}=-100 $ V) and 4-probe field-effect mobility extracted from the four-probe conductance in the high gate-voltage linear region [$\mu_{\mathrm{FE-4P}}=(1/C_{\mathrm{Geo}})\times(d\sigma/dV_{\mathrm{bg}})$].  At $T > 100$ K, the Hall and field-effect mobilities nearly overlap, and follow a power-law dependence, $ \mu\sim T^{-\gamma} $ with $ \gamma\approx 1.5 $ in good agreement with optical-phonon limited mobility~\cite{Xu2016, Chuang2016, Radisavljevic2013, Kaasbjerg2012}.  As temperature decreases, both the Hall mobility and field-effect mobility start to deviate from the power-law dependence, which may be attributed to the combination of acoustic phonon scattering and charged impurity scattering. In addition, the discrepancy between the Hall and field-effect mobilities also increases with decreasing temperature. At 10~K and 20~K, the field-effect mobility values are significantly higher than the Hall mobility values,  suggesting that field-effect mobility of the WSe$_2$ device may be overestimated at low temperatures, which has been previously explained by the carrier-density dependence of the actual mobility~\cite{Baugher2013}. In the gate-voltage range where the actual mobility is independent of carrier density, the field-effect mobility should be equal to the actual mobility. On the other hand, if the mobility increases with carrier density, the field-effect mobility is expected to overestimate the mobility~\cite{Baugher2013}. To verify this scenario, the Hall mobility is plotted as a function of carrier density for temperatures between 10 and 300 K, as shown in Figure~\ref{fig2}(c). At $T > 80$ K, the Hall mobility is nearly independent of carrier density, which explains the agreement between the field-effect mobility and Hall mobility in the relatively high temperature-range. As temperature decreases, the Hall mobility becomes increasingly carrier-density dependent. At $T = 10$~K, the Hall mobility increases from $\sim 1750$ cm$^2$V$^{-1}$s$^{-1}$ at $\sim2.7 \times10^{12}$ cm$^{-2}$ to $\sim2300$ cm$^2$V$^{-1}$s$^{-1}$ at $\sim4.5 \times10^{12}$ cm$^{-2}$, indicating that the mobility of the device is limited by charged impurity scattering at low temperature. Increased carrier density enhances screening of the Coulomb potential from the charged impurities, leading to improved mobility. 

Investigation of the 2D MIT and the associated dynamical scaling typically requires temperatures that are lower than the Fermi energy $E_\mathrm{F}$ of the 2D system  \cite{Dragana2017}.  In Si MOSFETs, in which the critical density for the MIT is $n_c(10^{11}$cm$^{-2})\sim0.5-5$ (Ref.~\citenum{Lin2015})
and the corresponding Fermi temperature $T_\mathrm{F}\sim4-35$~K  ($T_\mathrm{F}$[K]$=7.31\,n_s(10^{11}$cm$^{-2})$; Ref.~\citenum{AFS}), this typically requires cryogenic temperatures of a $^3$He system or a dilution refrigerator.  Therefore, measurements made on Si MOSFETs in these temperature ranges generally use drain-source voltages $V_{\mathrm{ds}}\sim5-100~\mu$V to avoid heating~\cite{Lin2015, Feng2001, Simmons1998, Popovic1997,Bogdanovich2002}.  For our WSe$_2$ system, $45 \leq T_\mathrm{F}$~[K]~$ \leq 220$ in the measurement range $-40$~V $\leq V_{\mathrm{bg}}\leq -90$~V (Ref.~\citenum{Movva2015}; see also Supporting Information), i.e. about an order of magnitude higher than in Si MOSFETs.  However, transport measurements on TMDs typically use $V_{\mathrm{ds}}$ that are several orders of magnitude larger~\cite{nihar2015, moon_soft_2018, Radisavljevic2013, Liu2013, Schmidt2014}. Even in the case of Ohmic contacts, this may cause substantial current heating and consequently lead to non-Ohmic behavior.  This suggests that a more nuanced approach to determining the Ohmic regime is needed when studying TMDs at low temperatures.

\begin{figure}[!t]
    \includegraphics[width=\linewidth]{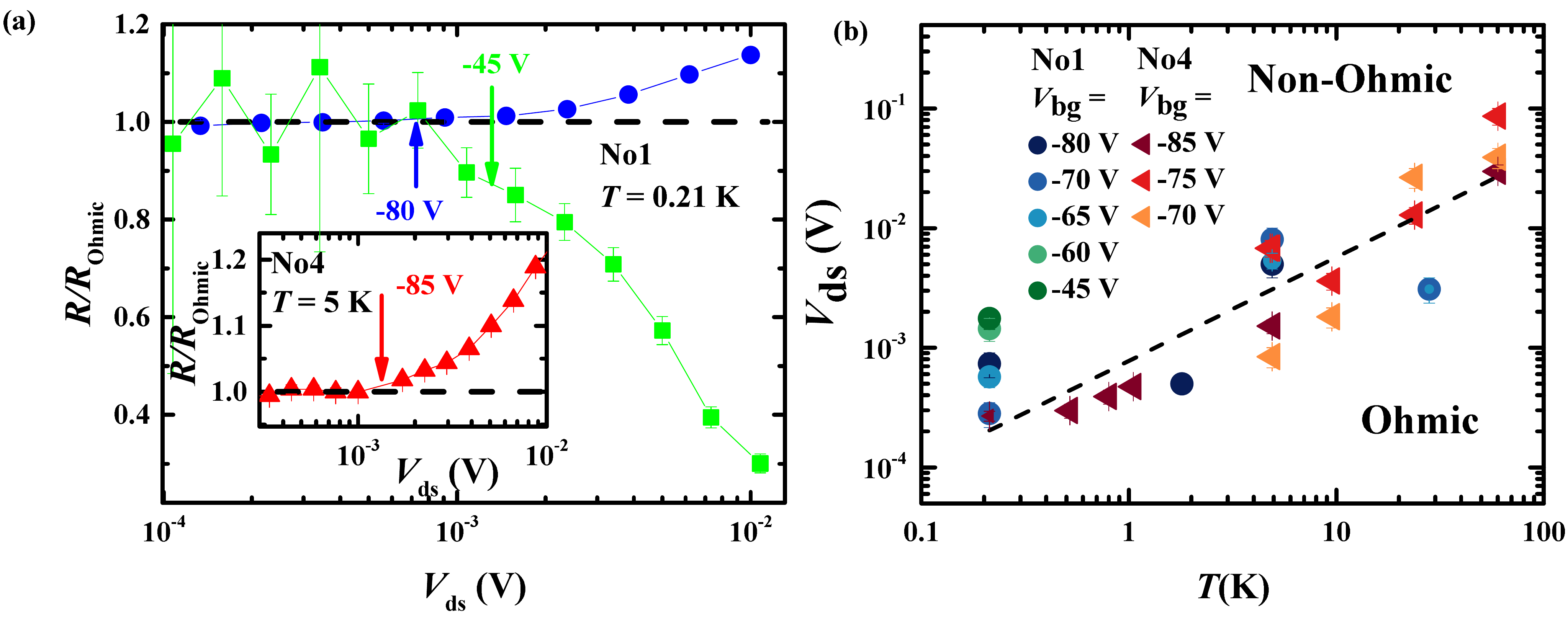}
  \caption{Determining the onset of non-Ohmicity.  (a) Normalized resistance as a function of drain-source voltage for sample No1 at 0.21 K and $V_{\mathrm{bg}}= -45$~V (green squares) and $-80$~V (blue dots), corresponding to the insulating and metallic regimes, respectively. The arrows indicate the corresponding onsets of non-Ohmicity, defined as the $V_{\mathrm{ds}}$ where $R$ deviates by 2 SD away from a constant value, $R_\mathrm{Ohmic}$; $R_\mathrm{Ohmic}$ was found by averaging the resistances measured at a few lowest $V_{\mathrm{ds}}$. Inset: The onset of non-Ohmic behavior for sample No4 at 5~K and $V_{\mathrm{bg}}=-85$~V in the metallic regime. (b) Onset of non-Ohmicity as a function of temperature for various back-gate voltages, as shown. Measurements on both samples No1 and No4 are included. The dashed line is a linear fit with a slope of $ 0.8\pm0.1 $.}
  \label{fig3}
\end{figure}

The standard method of checking the Ohmicity of the sample is to measure the current-voltage ($I_{\mathrm{ds}}$--$V_{\mathrm{ds}}$) characteristics, and determine whether the $I_{\mathrm{ds}}$--$V_{\mathrm{ds}}$ curves, when plotted on a linear scale, are linear in the limit of low excitations.  We have demonstrated in our previous study of WSe$_2$ FETs with 2D/2D contacts that the output characteristics measured at low biases remain linear down to cryogenic temperatures \cite{Chuang2016}.  At high enough excitations, of course, deviations from Ohmic behavior are always expected because of the heating. Using this method, however, it may be difficult to determine the onset of nonlinearity precisely.  Therefore, we employ a more sensitive technique, which involves measuring the deviation of the \textit{four-terminal } sample resistance $R$ from its constant, Ohmic value, as a function of $V_{\mathrm{ds}}$ to determine the Ohmic regime for our subsequent investigation of the 2D MIT.  In particular, the samples were held at a fixed $T$ and back-gate voltage $V_{\mathrm{bg}}$, while the $V_{\mathrm{ds}}$ was swept and current measured.  The onset of non-Ohmic behavior, $V_{\mathrm{ds}\mathrm{(onset)}}$, is determined as the $V_{\mathrm{ds}}$ at which $R$ starts to deviate from a constant, Ohmic value found at the lowest excitations [Figure~\ref{fig3}(a)].

\begin{figure}[t!]
\centering
      \includegraphics[width=\linewidth]{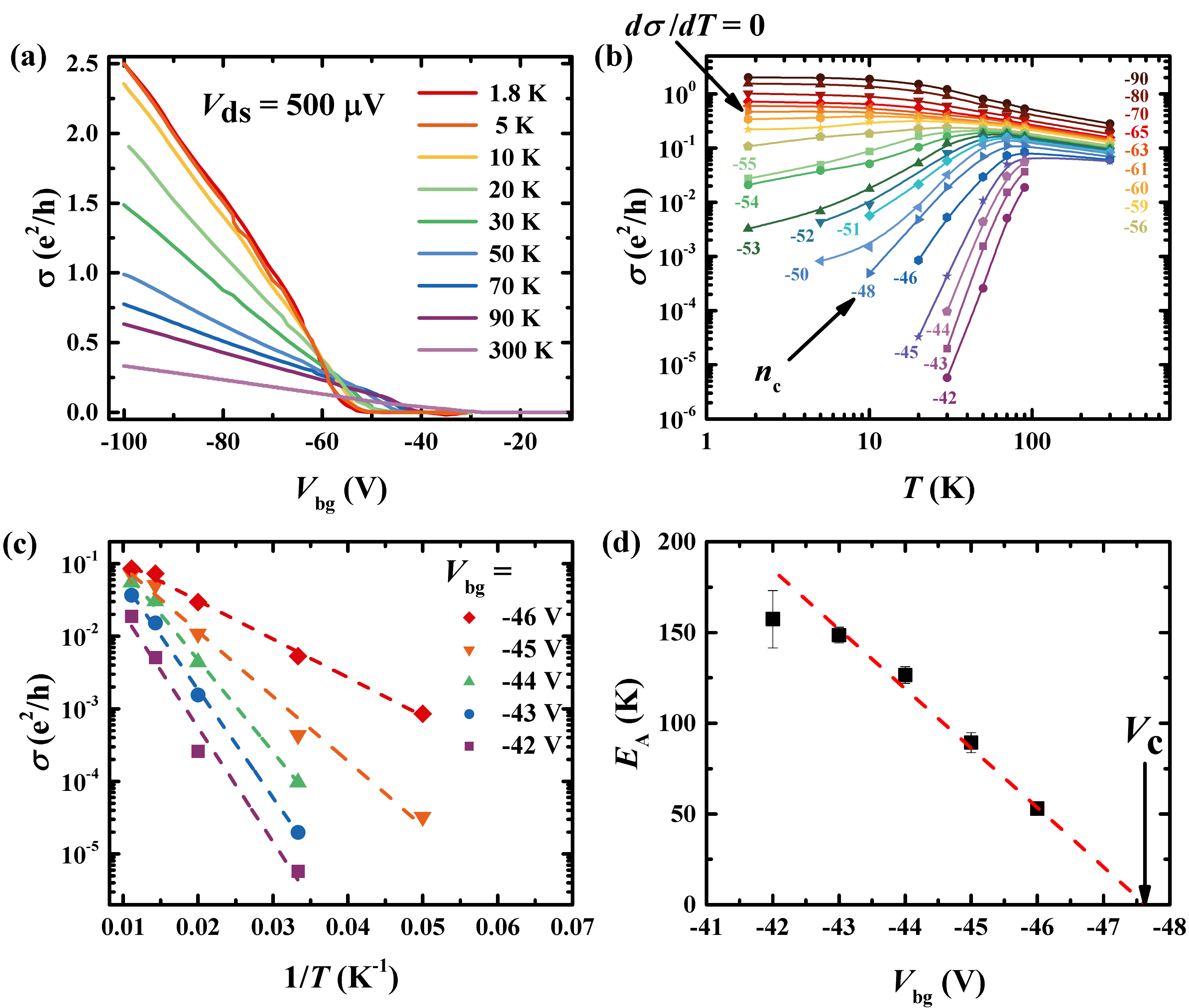}
  \caption{(a) Four-terminal conductivity $\sigma$ of sample No1 as a function of $V_{\mathrm{bg}}$ at various temperatures. (b) Temperature dependence of the conductivity $\sigma$ for various $V_{\mathrm{bg}}$[V], as shown. These data were extracted from (a). An arrow shows the critical density for the MIT, $n_\mathrm{c} = (1.29\pm0.03)\times10^{12}$ cm$^{-2}$.  Another arrow points at the density $n_{s}^{\ast}\approx 2.1\times10^{12}$ cm$^{-2}$ at which $d\sigma/dT=0$. (c) Conductivity as a function of inverse temperature for several $V_{\mathrm{bg}}$ on the insulating side of the MIT. Dashed lines are fits to $\sigma\propto\exp(-E_\mathrm{A}/T)$.  (d) Activation energy $E_\mathrm{A}$ as a function of $V_{\mathrm{bg}}$. The linear fit is shown by the dashed line and an arrow indicates the critical voltage $V_\mathrm{c}$, i.e. the back-gate voltage at the critical density for the MIT; $V_\mathrm{c}= -47.6 \pm 0.5 $~V.}
  \label{fig4}
\end{figure}

The deviation of the $R(V_{\mathrm{bg}})$ values from the corresponding $R_\mathrm{Ohmic}(V_{\mathrm{bg}})$, which is observed at high enough $V_{\mathrm{ds}}$, is consistent with heating: $R$ either increases or decreases with $V_{\mathrm{ds}}$, depending on the sign of its temperature dependence for a given $V_{\mathrm{bg}}$.  For example, $V_{\mathrm{bg}}=-45$~V corresponds to the insulating regime [Figure~\ref{fig4}(a)], i.e. $dR/dT<0$ ($d\sigma/dT>0$) at the lowest $T$, so that heating will lead to a decrease in the sample resistance, as seen in Figure~\ref{fig3}(a).  The opposite is observed for $V_{\mathrm{bg}}=-80$~V and $V_{\mathrm{bg}}=-85$~V (Figure~\ref{fig3}(a) inset), which are in the metallic regime with $dR/dT>0$ ($d\sigma/dT<0$).

Figure \ref{fig3}(b) summarizes the results for  $V_{\mathrm{ds}\mathrm{(onset)}}$ as a function of temperature for both samples. On a log-log scale, they follow a linear trend with the slope of $ 0.8\pm0.1$, thus further indicating that the $V_{\mathrm{ds}\mathrm{(onset)}}$ is determined primarily by temperature, i.e. by heating effects, and not the applied $V_{\mathrm{bg}}$.  It is clear that, similar to Si MOSFETs, at most $V_{\mathrm{ds}} \sim100~\mu$V may be used to be in the Ohmic regime at the lowest $T\sim 0.25$~K.  In fact, even at 100~K, $V_{\mathrm{ds}}=100$~mV typically used to study the 2D MIT in TMDs \cite{moon_soft_2018, nihar2015} puts the sample into the non-Ohmic regime due to heating.  The low-temperature Ohmic contacts in our WSe$_2$ Hall-bar devices enable us to measure the conductivity for wide ranges of temperature and carrier density at low enough excitation voltages to avoid heating effects. 

Using a $V_{\mathrm{ds}}$ in the Ohmic regime, we carried out measurements of the conductivity as a function of gate voltage, $\sigma(V_{\mathrm{bg}})$, at fixed temperatures [Figure~\ref{fig4}(a)].  It is apparent that $\sigma(T)$ changes its behavior from insulating at the lowest $V_{\mathrm{bg}}$ to metallic at the highest $V_{\mathrm{bg}}$. To better understand the temperature dependence and identify different regimes, we have extracted $\sigma(n_s, T)$ curves at fixed $V_{\mathrm{bg}}$ [Figure~\ref{fig4}(b)]. It is clear that, at low temperatures, the slope $ d\sigma/dT$ changes sign for $V_{\mathrm{bg}}\approx -61$~V.  This is sometimes attributed to the metal-insulator transition \cite{Radisavljevic2013}.  However, we note that, in general, $ d\sigma/dT>0 $ does not necessarily imply the existence of an insulating state, i.e. $\sigma(T=0)=0$. In fact, studies of the 2D MIT in Si MOSFETs have seen the existence of a 2D metal with $d\sigma/dT>0$ in low-disorder samples in the presence of local magnetic moments \cite{Feng2001, Eng2002}, as well as in high-disorder samples \cite{Bogdanovich2002,Lin2015}. Moreover, similar behavior is well-known in conventional semiconductors such as doped silicon, where the 3D metal-insulator transition occurs between a metallic state with $d\sigma/dT>0$ and $\sigma(T=0)\neq0$, and an insulating state with $d\sigma/dT>0$ and $\sigma(T=0)=0$ \cite{Sarachik1995}.  Therefore, in all those cases the critical density $n_{\mathrm{c}}$ for the MIT is lower than the density $n_{s}^{\ast}$ at which $d\sigma/dT=0$.

The critical carrier density $n_\mathrm{c}$, i.e. the critical back-gate voltage $ V_\mathrm{c} $, can be found instead from the fits to $\sigma(n_s,T)$.  In particular, on the insulating side of the transition, i.e. at the lowest $V_{\mathrm{bg}}$, we find that the data are best described by an activated temperature dependence, $\sigma\propto\exp(-E_\mathrm{A}/T)$ [Figure~\ref{fig4}(c)]. The vanishing of the activation energy $ E_\mathrm{A} $, i.e. the vanishing of strong localization, with increasing carrier density is often used as a criterion to determine $n_\mathrm{c}$ \cite{Lin2015, Bogdanovich2002, Pudalov1993, Shashkin2001, Jaroszynski2002,Jaroszynski2004}. Figure~\ref{fig4}(d) shows that the activation energy vanishes at $ V_\mathrm{c} = -47.6 \pm 0.5 $~V, which is obviously much lower than $V_{\mathrm{bg}}\approx -61$~V (carrier density $n_{s}^{\ast}\approx 2.1\times10^{12}$ cm$^{-2}$) where $d\sigma/dT $ changes sign [see Figure~\ref{fig4}(b)].  The critical $ V_\mathrm{c} = -47.6 \pm 0.5 $~V  corresponds to the critical density $n_\mathrm{c} = (1.29\pm0.03)\times10^{12}$ cm$^{-2}$; $T_{\mathrm{F}}(n_s=n_\mathrm{c})\approx 72$~K.  While this $n_\mathrm{c}$ is approximately an order of magnitude higher than in Si MOSFETs, it is comparable to or lower than what has been observed in MoS$_2$ and ReS$_2$, where $n_\mathrm{c}\approx 3\times10^{12}$ cm$^{-2}$ \cite{nihar2015, moon_soft_2018}.  The strength of the Coulomb interactions between charge carriers is usually characterized by the dimensionless Wigner-Seitz radius $r_s$, the ratio of the average Coulomb energy per carrier and $E_\mathrm{F}$.  In WSe$_2$, we find that $r_s\approx 5$ at the critical density (see Supporting Information).  A similar value of $r_s$ was reported in MoS$_2$ \cite{Radisavljevic2013}.  For comparison, in Si MOSFETs, the 2D MIT has been studied in samples with $4\lesssim r_s \lesssim 12$ at $n_s=n_\mathrm{c}$.

\begin{figure}[!tb]
      \includegraphics[width=\linewidth]{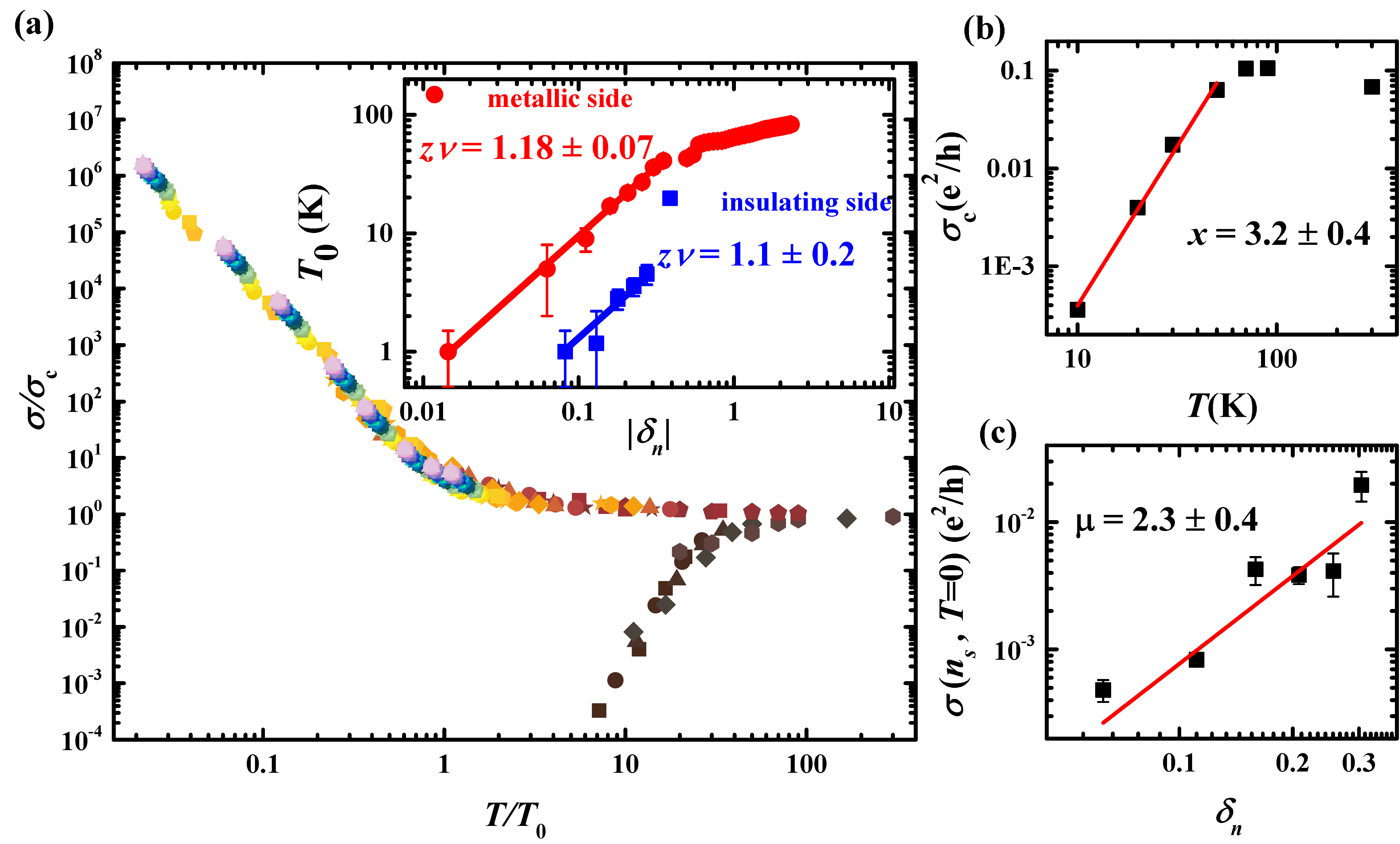}
  \caption{(a) Scaling of $ \sigma/\sigma_\mathrm{c }$ with temperature. Different colors and symbols represent $V_{\mathrm{bg}}$ from -42~V to -96~V (see Figure S2 in the Supporting Information for detailed labels).  The inset shows the scaling parameter $T_0$ vs $\delta_n$. Lines are fits with slopes $ z\nu = 1.1\pm0.2 $ and $ z\nu = 1.18\pm0.07 $ on the insulating and metallic sides, respectively.  (b) Temperature dependence of the critical conductivity $\sigma_\mathrm{c}=\sigma(V_{\mathrm{bg}} = V_\mathrm{c})$. The slope of the linear fit (red line) is the critical exponent $x = 3.2 \pm 0.4$. (c)  Zero-temperature conductivity on the metallic side of the MIT as a function of $ \delta_n $. The data were obtained by extrapolating the temperature dependent conductivity at fixed $V_{\mathrm{bg}}$ to $T=0$. The red line is a linear fit with the slope equal to the critical exponent $ \mu=2.3\pm0.4  $.}
  \label{fig5}
\end{figure}

Based on very general arguments, near the metal-insulator quantum phase transition the conductivity is expected  \cite{Belitz1994} to follow a scaling form $ \sigma(n_s,T)= \sigma_\mathrm{c}(T) f(T/T_0)$, where the critical conductivity $ \sigma_\mathrm{c}(T)=\sigma(n_s=n_\mathrm{c},T)\propto T^x $.  The crossover temperature $T_0\propto \delta^{z\nu}_n$, where $z$ and $\nu$ are the dynamical and correlation length exponents, respectively, and $ \delta_n = (n_s -n_\mathrm{c})/n_\mathrm{c} $.  Figure~\ref{fig5} (a) shows that, near $n_\mathrm{c}= (1.29\pm0.03)\times10^{12}$ cm$^{-2}$, 
it is indeed possible to collapse all $ \sigma(n_s,T)/\sigma_\mathrm{c}(T) $ onto the same function $f(T/T_0)$, with two branches.  The scaling is achieved over a large, four-orders-of-magnitude range of $T/T_0$.  The upper branch corresponds to the metallic side of the transition and the lower branch to the insulating side. The scaling parameter $T_0$ follows the same power-law function $T_0 \propto |\delta_n|^{z\nu}$ on both sides of the transition (Figure~\ref{fig5}(a) inset), within experimental error, as expected for a quantum phase transition \cite{Belitz1994}; $ z\nu = 1.1\pm0.2 $.  Figure~\ref{fig5}(b) shows that $ \sigma(n_\mathrm{c},T) \propto T^x $ with $x = 3.2 \pm 0.4$.  We note that, surprisingly, in Figure~\ref{fig5}(a) it was possible to scale the data even up to fairly high $T>T_{\mathrm{F}}$ for densities very close to $n_\mathrm{c}$.  If only the lower-temperature data, such that $T<T_\mathrm{F}$, are included in scaling, the critical exponents remain the same within error (see Figure~S3 in the Supporting Information).

Furthermore, on the metallic side of the transition, a power-law behavior is expected for the zero-temperature conductivity, $ \sigma(n_s, T=0)\propto \delta_n^{\mu}$ \cite{Belitz1994}.  By extrapolating the $ \sigma(n_s, T)$ curves to $T=0$, we have determined $ \sigma(n_s, T=0)$, and found that they obey $ \sigma(n_s, T=0)\propto \delta_n^{\mu}$ with $ \mu = 2.3 \pm 0.7$ [Figure~\ref{fig5}(c)].  Notably, the extrapolated $ \sigma(n_s, T=0)$ values go to zero at the same $n_\mathrm{c}$ where activation energies, determined from the insulating side, vanish, thus further confirming the determination of $n_\mathrm{c}$.

Finally, from standard scaling arguments \cite{Belitz1994}, it follows that the critical exponent $\mu$ can be determined not only from extrapolations of $\sigma(n_s,T)$ to $T=0$, but also from $ \mu =x(z\nu)$ based on all data taken at all $T$ and $n_s$ for which scaling holds.  Indeed, using $x = 3.2 \pm 0.4$ [Figure~\ref{fig5}(b)] and $ z\nu = 1.1\pm0.2 $ [Figure~\ref{fig5}(a)], we find $ \mu =x(z\nu) = 3.6 \pm 0.8$, which agrees within experimental error with $ \mu = 2.3 \pm 0.7$ found from $T=0$ extrapolation of $\sigma(n_s, T)$ [Figure~\ref{fig5}(c)].  This confirms the consistency of the analysis.  

The high quality of our devices with Ohmic contacts has enabled us to perform the measurements down to low $T$, with low enough excitations to avoid heating effects and, in particular, to reliably measure into the insulating side of the MIT.  This has allowed us to determine the values of $z\nu$ that are equal for both branches of the scaling function, something that has been difficult to achieve in other TMDs \cite{nihar2015,moon_soft_2018}.  In addition, since we have been able to determine the carrier density as a function of $V_{\mathrm{bg}}$ accurately, the values of our critical exponents are not subject to a large uncertainty caused by the differences in the carrier densities derived from Hall measurements and the parallel-plate capacitor model \cite{nihar2015}.  While the critical exponents for MoS$_2$ are still a subject of investigation \cite{moon_soft_2018,Moon2019}, our critical exponents for WSe$_2$, like those for ReS$_2$ \cite{nihar2015}, are consistent, within error, with those seen in Si MOSFETs with low disorder in the presence of local magnetic moments~\cite{Feng2001}, one of the proposed universality classes \cite{Lin2015} for the 2D MIT.  In Si MOSFETs, the origin of local moments was attributed to singly occupied localized states in the tail of an upper electric subband, which then act as additional scattering centers for 2D electrons.  By using the top gate and the back gate, it was possible to vary the subband splitting and thus the population in the tail of the upper subband in a controlled way, while keeping the 2D density $n_s$ fixed \cite{Feng1999}.  It is interesting to speculate whether similar multiband effects might be responsible also for $\sigma(n_s, T)$ observed in ReS$_2$ (Ref.~\citenum{nihar2015}) and here in WSe$_2$, and in particular, for their  relatively large activation energies $E_\mathrm{A} \gtrsim E_\mathrm{F}$ in the insulating regime (Ref.~\citenum{nihar2015} and Figure~\ref{fig4}(d), respectively).  Perhaps this could be tested by using TMD-based double-gate FETs, in analogy to Si MOSFET studies.

\section{Conclusions}

Our study on WSe$_2$ has demonstrated that 2D/2D Ohmic contacts technology described in Ref.~\citenum{Chuang2016} represents an exciting and reliable platform to investigate the 2D MIT in new TMD-based materials. In conjunction with hBN encapsulation, this contact-engineering technique produces high-quality Hall-bar devices with low-temperature Ohmic contacts down to 0.25~K. This has allowed us to demonstrate scaling behavior consistent with the existence of a 2D metal-insulator quantum phase transition.  The critical exponents are found to agree with those seen in low-disorder Si MOSFETs, in which the 2D MIT is driven by electron-electron interactions, but disorder-induced local magnetic moments are also present.  Our results thus support the universality of the behaviors observed in Si MOSFETs, and suggest that similar mechanisms might be at play also in WSe$_2$.  Further insight into the interplay of interactions and disorder in 2D systems, and into the microscopic origin of the MIT, could be gained from studies of quantum criticality in other TMD-based materials using the same contact engineering.

\begin{suppinfo}
\begin{itemize}
  \item Supporting Information: Charge density and mobility of samples No2 and No3, calculation of the Fermi energy and $r_s$, additional details about the 
  scaling of conductivity
\end{itemize}

\end{suppinfo}
\vspace*{24pt}

\noindent\textbf{\Large{Author information}}

\noindent\textbf{Corresponding Authors}

\noindent*E-mail: zxzhou@wayne.edu (Z. Z.)

\noindent*E-mail: dragana@magnet.fsu.edu (D. P.)

\noindent \textbf{Author contributions}

\noindent M.~K., J.~Y. and D.~M. worked on synthesis and basic characterization of bulk crystals.  H.-J.~C. fabricated Hall-bar devices.  L.~J.~S. and H.-J.~C. performed the electrical measurements.  L.~J.~S., H.-J.~C. and Z.~Z. analyzed the data.  L.~J.~S., D.~P., and Z.~Z. discussed the results and wrote the manuscript, with input from all authors.  D.~P. and Z.~Z. supervised the project.

\noindent\textbf{Notes}

\noindent The authors declare no competing financial interest.

\begin{acknowledgement}

The work by L.~J.~S. and D.~P. was supported by NSF Grants Nos. DMR-1307075 and DMR-1707785, and the National High Magnetic Field Laboratory through the NSF Cooperative Agreement No. DMR-1644779, and the State of Florida.  Z.~Z. acknowledges support by NSF grant No. DMR-2004445.  J.~Y. and D.~M. acknowledge support from the U.S. Department of Energy, Office of Science, Basic Energy Sciences, Materials Sciences and Engineering Division. 

\end{acknowledgement}


\clearpage

\noindent\textbf{\Large{Supporting Information}}

\makeatletter
\makeatletter \renewcommand{\fnum@figure}{{\bf{\figurename~S\thefigure}}}
\makeatother

\setcounter{figure}{0}
\setcounter{section}{0}

\baselineskip=24pt

\section{Charge density and mobility of samples No2 and No3}

\begin{figure}[ht!]
\centering
    \includegraphics[width=0.8\textwidth]{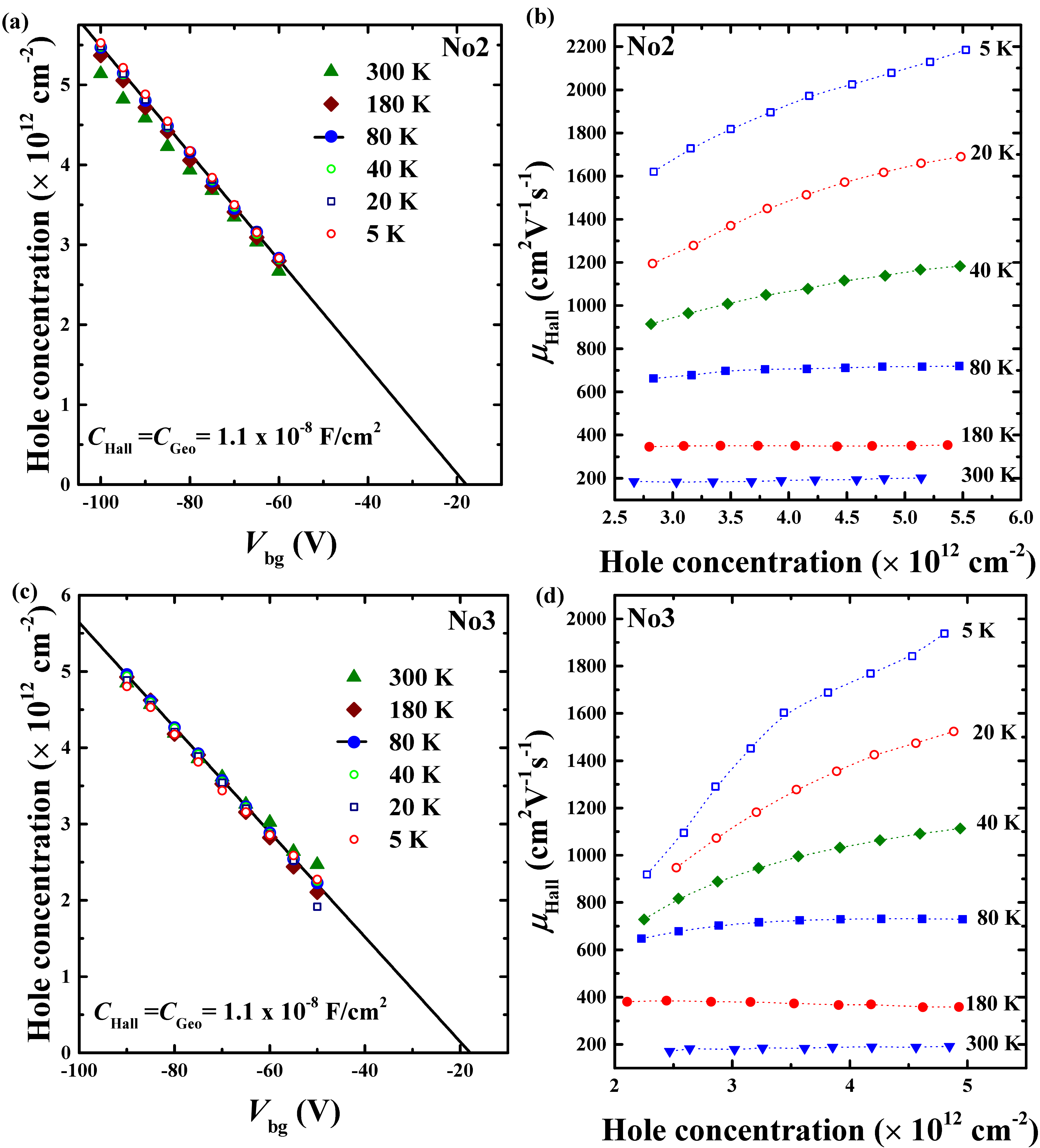}
  \
      \caption{(a) Hole concentration determined from Hall measurements, $n_{\mathrm{Hall}}$, as a function of back-gate voltage for multiple temperatures, and (b) Hall mobility as a function of hole density for selected temperatures, for sample No2 (thickness $\approx 4.9$~nm). (c) Hole concentration  $n_{\mathrm{Hall}}$ as a function of back-gate voltage for multiple temperatures, and (d) Hall mobility as a function of hole density for selected temperatures, for sample No3 (thickness $\approx 7.9$~nm).  Solid lines in (a) and (c) are linear fits of the 80~K data. Dashed lines in (b) and (d) guide the eye. 
      }
  \label{figS1}
\end{figure}

Figure~\ref{figS1} shows carrier concentration $n_{\mathrm{Hall}}$ vs. gate voltage and Hall mobility vs. carrier density for two more devices with different WSe$_2$ thicknesses. Both devices exhibit qualitatively and quantitatively similar behavior as the WSe$_2$ device shown in Figure 2.  

\section{Calculation of the Fermi energy and $\bm{r_s}$}

The Fermi energy $E_\mathrm{F}$ is given by 

\begin{equation}
k_BT_\mathrm{F}=E_\mathrm{F}=\frac{\hbar^2 k_\mathrm{F}^2}{2m^{\ast}}=\frac{\pi\hbar^2n_s}{m^{\ast}},
\end{equation} 
where $k_B$ is the Boltzmann constant, $\hbar$ is the reduced Planck constant, $T_\mathrm{F}$ is the Fermi temperature, $k_\mathrm{F} =\sqrt{2\pi n_s}$ is the Fermi wave vector, $n_s$ is the carrier density, and $m^{\ast}$ is the hole effective mass.  We assume $m^{\ast}=0.5m_e$ ($m_e$ is the free electron mass), a valley degeneracy $g_v=1$ and a spin degeneracy of 2 to relate $k_\mathrm{F}$ to $n_s$~[Ref.~(1)].

We can characterize the strength of carrier interactions using the dimensionless Wigner-Seitz radius, 
\begin{equation}
r_s=\frac{E_\mathrm{C}}{E_\mathrm{F}}=\frac{1}{\sqrt{\pi n_s}a_B}= \frac{m^{\ast}q^2g_v}{4\pi\varepsilon\varepsilon_0\hbar^2 \sqrt{\pi n_{s}}},
\end{equation}
where $E_\mathrm{C}$ is the average Coulomb energy per charge carrier, $a_B$ is the Bohr radius in a semiconductor, $q$ is the charge of the electron, 
$ \varepsilon\approx 10$ [Ref.~(2)]  
is the relative dielectric constant of WSe$_2$, and $ \varepsilon_0$ is the vacuum dielectric constant. 

For our sample No1 we found that $n_\mathrm{c} =(1.29\pm0.03)\times10^{12}$ cm$^{-2}$, thus we calculate that at the critical density $T_\mathrm{F}= 72$ K and $r_s \approx 5$.

\clearpage

\section{Scaling of conductivity}

\begin{figure}[!ht]
\centering
    \includegraphics[width=0.7\linewidth]{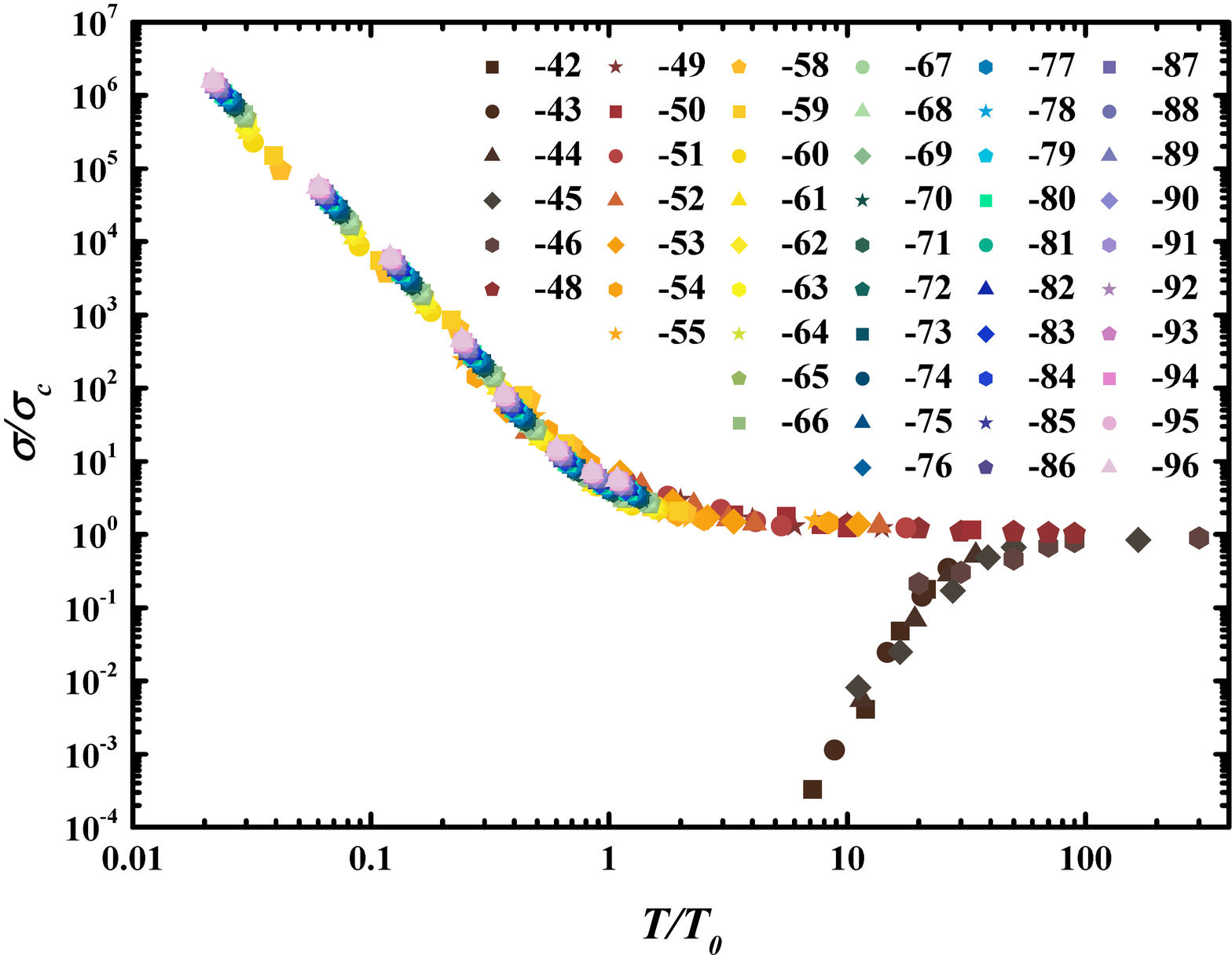}
    \vspace*{-12pt}
  \caption{Scaling of $ \sigma/\sigma_\mathrm{c} $ with temperature shown in Figure 5(a). Different colors and symbols correspond to different $V_{\mathrm{bg}}$[V], as shown.}
  \label{fig-scaling-labels}
\end{figure}

\begin{figure}[!ht]
\centering
    \includegraphics[width=\linewidth]{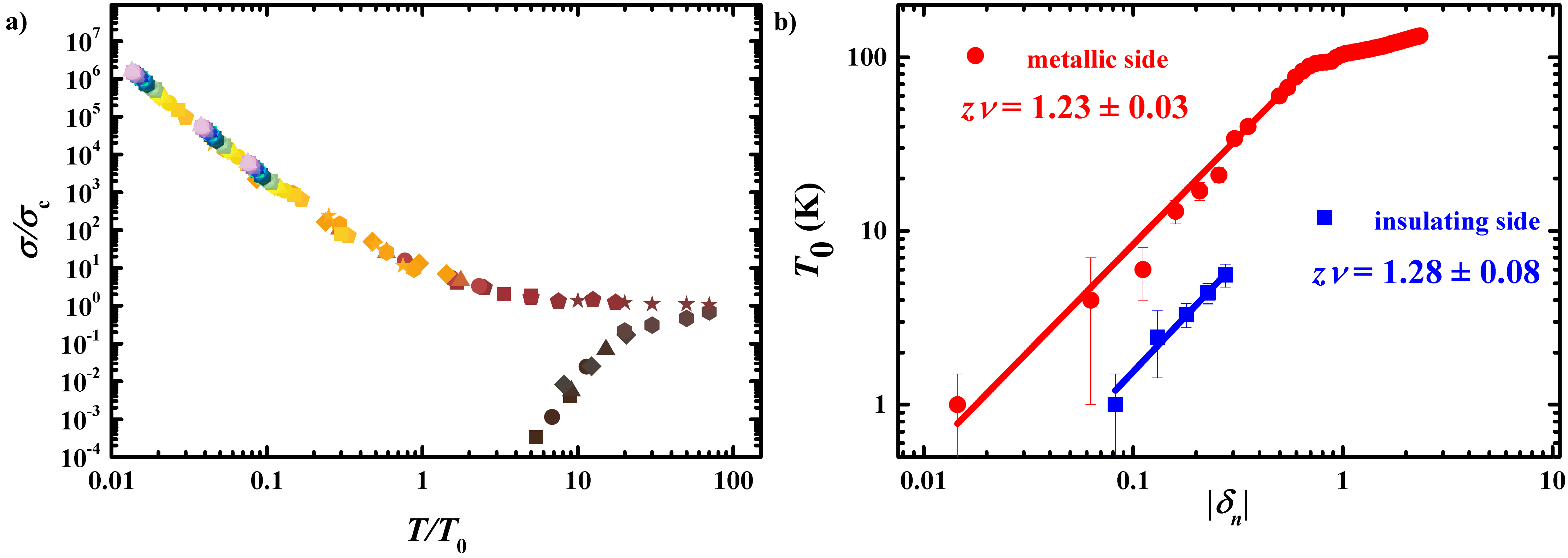}
  \caption{Scaling with only the $T<T_\mathrm{F}$ data.  (a) Scaling of $ \sigma/\sigma_\mathrm{c} $ with temperature. Different colors and symbols represent $V_{\mathrm{bg}}$ from -42~V to -96~V.  The inset shows the scaling parameter $T_0$ vs $\delta_n$. Lines are fits with slopes $ z\nu = 1.28\pm0.08 $ and $ z\nu = 1.23\pm0.03 $ on the insulating and metallic sides, respectively.}
  \label{figS2}
\end{figure}


\clearpage

\noindent\textbf{Supporting Information References}

\noindent (1) Movva, H. C. P.; Rai, A.; Kang, S.; Kim, K.; Fallahazad, B.; Taniguchi, T.; Watanabe, K.; Tutuc, E.; Banerjee, S. K. High-Mobility Holes in Dual-Gated WSe$_2$ Field-Effect Transistors. \textit{ACS Nano} \textbf{2015}, \textit{9}, 10402--10410.

\noindent (2) Kang, Y,; Jeon, D.; Kim, T. Direct Observation of the Thickness-Dependent Dielectric Response of MoS$_2$ and WSe$_2$.  \textit{J. Phys. Chem. C } \textbf{2020}, \textit{124}, 18316--18320.

\end{document}